\documentclass[letterpaper,12pt]{article}
\usepackage{setspace}
\usepackage{tabu}
\usepackage{times}
\usepackage{booktabs}
\usepackage[table,xcdraw]{xcolor}
\usepackage{charter}
\usepackage[english]{babel}
\usepackage{wrapfig}
\usepackage{upgreek}
\usepackage{siunitx}
\usepackage[font=small,labelfont=bf]{caption}
\usepackage{titlesec}
\usepackage{graphicx}
\usepackage{pgfplots}
\usepackage{fancyhdr} 

 \newcommand{\parallelsum}{\mathbin{\!/\mkern-5mu/\!}}
\titleformat{\section}
  {\normalfont\fontsize{11}{12}\bfseries}{\thesection}{1em}{}
\titleformat{\subsection}
  {\normalfont\fontsize{11}{12}\bfseries}{\thesubsection}{1em}{}
\titleformat{\subsubsection}
  {\normalfont\fontsize{11}{12}\bfseries}{\thesubsubsection}{1em}{}
  
\titlespacing*\section{0pt}{3mm plus 1pt minus 2pt}{1mm plus 2pt minus 2pt}
\titlespacing*\subsection{0pt}{3mm plus 1pt minus 2pt}{1mm plus 2pt minus 2pt}
\titlespacing*\subsubsection{0pt}{3pt plus 3pt minus 2pt}{0pt plus 2pt minus 2pt}
\titlespacing*\paragraph{0pt}{6pt plus 3pt minus 2pt}{0pt plus 2pt minus 2pt}

\graphicspath{ {./Figures/} }

\usepackage[letterpaper,top=1in,bottom=1in,left=1in,right=1in]{geometry}

\usepackage{amsmath}
\usepackage{graphicx}
\usepackage[colorinlistoftodos]{todonotes}
\usepackage{siunitx}
\usepackage{latexsym}
\DeclareSIUnit{\sqrthz}{\ensuremath{\sqrt{\text{\hertz}}}}
\pgfplotsset{compat = 1.14}

\title{Design and Characterization of Superconducting Nanowire-Based Processors for Acceleration of Deep Neural Network Training}
\author{\small Murat Onen$^1$, Brenden A. Butters$^1$, Emily Toomey$^1$, Tayfun Gokmen$^2$, Karl K. Berggren$^1$}

\begin{document}

\date{ \small $^1$ Department of Electrical Engineering and Computer Science, Massachusetts Institute of Technology, Cambridge, Massachusetts 02139, USA\\ $^2$ IBM T.J. Yorktown Research Center, Yorktown Heights, NY, 10598, USA\\}
\maketitle

\setcounter{page}{1}
\noindent\textbf{Abstract:}

Training of deep neural networks (DNNs) is a computationally intensive task and requires massive volumes of data transfer. Performing these operations with the conventional von Neumann architectures creates unmanageable time and power costs. Recent studies have shown that mixed-signal designs involving crossbar architectures are capable of achieving acceleration factors as high as 30,000$\times$ over the state of the art digital processors. These approaches involve utilization of non-volatile memory (NVM) elements as local processors. However, no technology has been developed to-date that can satisfy the strict device requirements for the unit cell. This paper presents the superconducting nanowire-based processing element as a cross-point device. The unit cell has many programmable non-volatile states that can be used to perform analog multiplication. Importantly, these states are intrinsically discrete due to quantization of flux, which provides symmetric switching characteristics. Operation of these devices in a crossbar is described and verified with electro-thermal circuit simulations. Finally, validation of the concept in an actual DNN training task is shown using an emulator.

\noindent\textbf{Keywords: } Neural network accelerators, mixed-signal computing, superconducting nanoelectronics, crossbar architecture.

\clearpage
\setcounter{page}{1}
\pagestyle{fancy}
\section{Introduction}

Deep neural networks (DNNs) \cite{lecun2015} are powerful tools that are widely used in image recognition \cite{krizhevsky2012}, natural language processing \cite{collobert2011} and big-data analytics \cite{najafabadi2015deep}. Training of these structures involves a high number of matrix multiplications and require massive volumes of data transfer. As a result, time and power costs become unmanageable for larger implementations, which limits the scalability of the approach. 

Acceleration of DNN training has attracted great attention in hardware research (i.e. neuromorphic engineering). These approaches aim to build non-volatile memory (NVM) based local processors to reduce the high volumes of data transfer suffered by conventional von Neumann architectures \cite{burr2017neuromorphic}. Implementations involving phase change memory (PCM) \cite{ambrogio2018equivalent}, resistive random access memory (RRAM), \cite{chi2016prime} and memristors \cite{prezioso2015training,kim2011functional} have recently emerged to realize mixed-signal accelerators.

The most common framework employed in these implementations is the crossbar architecture \cite{steinbuch1961lernmatrix}. The main idea behind this method is to store the matrix entries (weights) in cross-point elements and perform multiplications locally. These frameworks reduce the $\mathcal{O}(N^3)$\footnote{Using textbook matrix multiplication. It can be optimized to $\mathcal{O}(\textit{$N^{2.373}$})$ using advanced algorithms \cite{coppersmith1987matrix}.} computational complexity of matrix-matrix multiplication for digital processors to $\mathcal{O}(N)$. Recently, an all-parallel update scheme has been demonstrated for crossbars that combines computation and application of the weight updates to obtain $\mathcal{O}(N)$ complexity as well \cite{gokmen2016}. This method extends the application of crossbars to training accelerators, on top of inference machines. However, it also requires a set of strict device properties for the unit cell that have not yet been achieved by any approaches to-date. Imperfections of the unit cells lead to heavy training performance degradation and limit the approach to inference applications only.

In this work, we first delineate the unit-cell requirements and then provide a superconducting nanowire-based design. Devices demonstrated in this paper have many programmable non-volatile states that can be used in analog multiplication. Furthermore, their switching characteristics are inherently symmetric due to the quantized nature of flux (equivalently, circulating current) in superconducting loops. The results shown in this paper suggest that this new family of devices can unlock the full potential of crossbar-based DNN training accelerators.

\section{Background: Crosspoint Element Requirements }

A crosspoint element can be defined as a local multiplier with a tunable multiplication factor. The number of its programmable states defines the resolution of the multiplier. A high ON/OFF ratio\footnote{Ratio of the highest multiplication factor to the lowest.} and a large number of states are desirable in order to have devices with superior controllability.
These states must also be non-volatile, with retention times longer than the complete training procedure in order to enable incremental and implicit updating. 

As previously emphasized by Ref.\cite{gokmen2017}, the switching dynamics between states should be symmetric with low inter-device variation. Here we define symmetry as the state of the cell remaining effectively unchanged, following an arbitrary sequence of increment and decrement pulses of equal total number. Considering that the DNN training consists of numerous small updates, the presence of any asymmetry creates drift terms in weight programming and leads to significant deterioration of the training performance\footnote{Some requirements with the crosspoint element can be relaxed by using simple algorithmic manipulations as explained in \cite{gokmen2017}. However, in a realistic scenario, all cells ($\textit{N}^2$) will bear different asymmetries. However, in a parallelized update scheme, only ($2\textit{N}$) variables can be accessed. This argument implies that the need for symmetry is absolute.}. Unlike asymmetric updates (that are of drift-like nature), random noise (of diffusion-like nature) in readout and/or update can be tolerated due to the intrinsic dynamics of neural network training \cite{gokmen2016,vincent2010stacked}. Finally, to realize the aforementioned stochastic update scheme, switching of the device requires a thresholding mechanism to allow state changing only under concurrent pulses on the row and column (see Sec. \ref{sec:operation}).

Due to the reasons listed above, to successfully implement an efficient DNN training accelerator, a symmetric switching multi-level programmable non-volatile unit cell is essential. These properties should be maintained at the device level since compensation techniques limit the scalability of the approach. Methods that involve serial accessing to individual devices reduce the acceleration factors significantly. On the other hand, approaches that introduce compliance circuits around the device impair the scalability of the approach with inefficient device counts. 

Upon the satisfaction of these strict criteria, mixed-signal frameworks can potentially accelerate DNN training 30.000$\times$ with respect to the state-of-the-art digital processors which is the reason behind the existing efforts in crosspoint element research in the nanoelectronics community \cite{gokmen2016}.
  
\section{Superconducting Nanowire-Based Processor}

In this section, we will first discuss the operation dynamics of the superconducting nanowire-based cross-point element. Then, a new method to perform analog multiplication with these devices will be explained. Finally, system-level details of the crossbar architecture will be shown with connection diagrams under different steps of the backpropagation algorithm, and we will present an analysis of the periphery requirements.

\subsection{State Representation and Programming}
\label{sec:SNIPE}

\begin{figure}[h!]
	\centering
	\includegraphics[width = \linewidth]{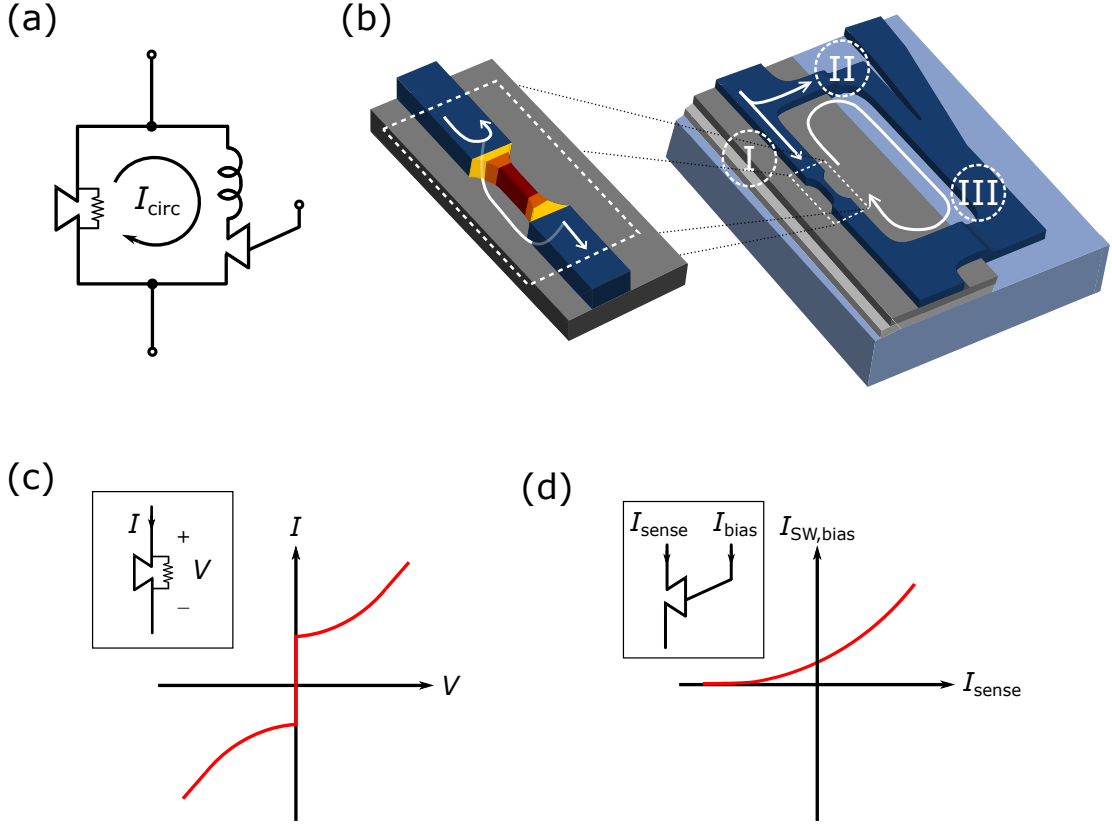}
	\caption{\textit{\textbf{Programming of the cell and the fundamental characteristics of the subcomponents:}} \textbf{a)} Circuit schematic of the unit cell. \textbf{b)}  Constriction switches into resistive state as a response to the increased current density on it (I). Due to the resistance of the constriction, current in the loop redistributes, which allows the constriction to heal back to its superconducting state (II). Once the loop is fully superconducting again, the excess current gets trapped inside the loop (III). \textbf{c)} Notional I-V curve of a shunted nanowire which shows non-hysteretic characteristics \cite{toomey2018bridging}. \textbf{d)} Notional characteristic graph for yTron, showing modulation of the switching current of the bias arm as a function of current flowing in the sense arm \cite{mccaughan2016}. Insets in \textbf{(c)} and \textbf{(d)} represent the circuit symbols for the shunted nanowire and yTron.}
	\label{fig:cartoon}
\end{figure}

The superconducting nanowire-based processing element is essentially a superconducting loop that employs a shunted constriction \cite{toomey2018bridging} as the writing element and a y-shaped current combiner (yTron) \cite{mccaughan2016} as the reading element (Fig.\ref{fig:cartoon}a). As the current flowing in the constriction branch is increased above nanowire's switching current ($I_\textrm{SW}$), it switches into resistive (normal) state  (Fig.\ref{fig:cartoon}b-I). This event redirects the current from the constriction branch into the loop (Fig.\ref{fig:cartoon}b-II). Under the absence of current flow, the nanowire restores its superconducting state and traps the excess current shuttled into the loop (Fig.\ref{fig:cartoon}b-III). The state of the unit cell can be represented by this circulating persistent current ($I_\textrm{circ}$) and can be controlled through the constriction.

For an unshunted nanowire, this switching event is abrupt, meaning that even a single event loads the loop to its maximum capacity. To have fine control over the current shuttled into the loop, the constriction can be locally shunted by a resistor \cite{toomey2018bridging}. This shunt branch dampens the switching characteristics of the constriction and prevents excess current from flooding into the loop all at once (i.e. a portion of the current gets redirected through the metal layer, Fig.\ref{fig:cartoon}b-II). If the impedance of the shunt branch is sufficiently low, this shuttled current can be equivalent to a single flux quantum (SFQ, $\Phi_\circ$) \footnote{Circulating currents in a superconducting loop can only occur in integer multiples of  $\Phi_\circ/L_\textrm{loop}$, where $\Phi_\circ = h/2e = \SI{2.07e-15}{\volt\s}$ is the flux quantum, and $L_\textrm{loop}$ is the device loop inductance. Therefore the finest control over the current is limited by this quantization.}\cite{toomey2018bridging}. Considering that there are no dissipative elements in the superconducting loop, this circulating current remains unchanged indefinitely, which means that the state is perfectly non-volatile. Furthermore, the inherently discretized nature of the system (flux quantization) ensures symmetric state switching under SFQ control, since there are no stable intermediate states.

The number of programmable states $N$ is given by  $ N = 2I_\textrm{SW}/\Delta I$, where $I_\textrm{SW}$ is the switching current\footnote{The factor 2 comes from the range of $\pm I_\textrm{SW}$} of the narrowest part of the loop (which is the constriction) and $\Delta I$ is the circulating current difference between adjacent states (equivalent to $L_\textrm{loop}\times\Phi_\circ$ under SFQ level control). Therefore, a higher number of programmable states can be achieved by increasing the loop inductance. Unlike normal metals, superconductors possess a high kinetic inductance, $L_\textrm{k}$ (arising from the inertia of the charge carriers \cite{annunziata2010tunable}). For such materials, the $L_\textrm{k}$ can be orders of magnitude higher than the geometric inductance. Therefore, large inductors can be realized without consuming excessive area. 

Finally, to read the state of these devices, the circulating current, $I_\textrm{circ}$, should be sensed. The current flowing in the sense arm modulates the switching current of the bias arm as illustrated by Fig.\ref{fig:cartoon}d. This effect is due to a simple geometric phenomenon known as the current crowding \cite{mccaughan2016}. Therefore, the point at which the bias arm switches indirectly measures the current flowing in the sense arm ($I_\textrm{circ}$ in a cell). Since the path of the circulating current does not experience any switching event, the state of the loop remains unchanged and the readout is non-destructive.

These features make this approach a perfect candidate for realizing a multi-level programmable non-volatile unit cell with inherently symmetric switching characteristics. 

\subsection{Analog Multiplication}
\label{sec:multiplication}
In a conventional crossbar with resistive cross-point elements, analog multiplication is implemented by Ohm's law where the weight is represented by the conductance of the element. Then, the contribution from each element is combined at the ends of columns via Kirchoff's law, forming the vector-matrix product. Superconducting devices, with infinite DC conductance, need a new approach to implement multiplication.

\begin{figure}[h!]
	\centering
	\includegraphics[width = 0.6\linewidth]{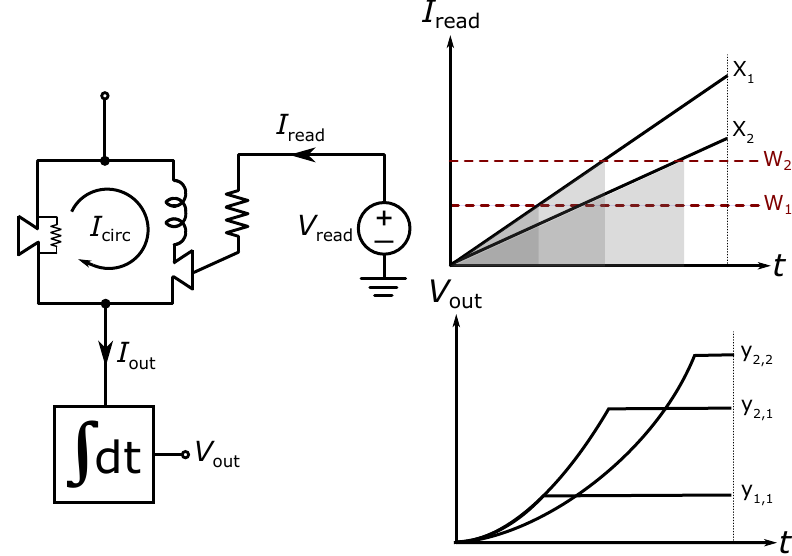}
	\caption{\textit{\textbf{Analog multiplication scheme for superconducting nanowire-based processor:}} The circuit schematic for the multiplication operation (left). Input currents ($I_\textrm{read}$) and output voltages ($V_\textrm{out}$) for three different cases (right). The state of the loop,  $W_\textrm{i}$, determines the integration time while the ramp height $x_\textrm{j}$ determines the integrand. The integrated output $y_\textrm{i,j}$ is proportional to $W^2x^{-1}$ for weight $W$ and input $x$.}
	\label{fig:Operations}
\end{figure}

The multiplication system adopted in this work involves representation of the input signals as voltage ramps (instead of constant voltage levels). 
This signal is then converted into current at each yTron (bias arm) with the use of a series resistor\footnote{Voltage ramps are provided across forward input lines(or backward input lines for the backward pass) where the other end of the yTron is connected to integrators, which are at ground level (Fig.\ref{fig:Crossbar}).}. As a response to this ramping input, the yTron's bias arm switches at a certain level that is determined by the state, $I_{\mathrm{circ}}$, of that unit cell (Fig.\ref{fig:Operations}, $\propto W/x$). After this switching event, the large induced resistance of the bias arm ($\sim$ \SI{}{\kilo\ohm}) results in a precipitous drop of the output current. The output of the multiplication operation is obtained by the integration of this output current. For a unit cell that is set to a higher state (larger $I_\textrm{circ}$), the bias arm switches at a higher current, corresponding to a later time in the ramps (see Fig.\ref{fig:cartoon}d), producing a larger integrated output. It can be analytically calculated that the integrated output is proportional to the product of the cell state and the input with the relation given as $W^2x^{-1}$.
\subsection{Crossbar Architecture and Operation}
\label{sec:operation}

The architecture used in this work is an implementation of the conventional crossbar approach \cite{steinbuch1961lernmatrix}, modified to be compatible with the stochastic update scheme. It is devised to support the training process with the stochastic gradient descent (SGD), using the backpropagation algorithm to compute the derivatives \cite{rumelhart1986learning} in all three operation cycles: forward pass, backward pass, and update.

\begin{figure}[ht]
	\centering
	\includegraphics{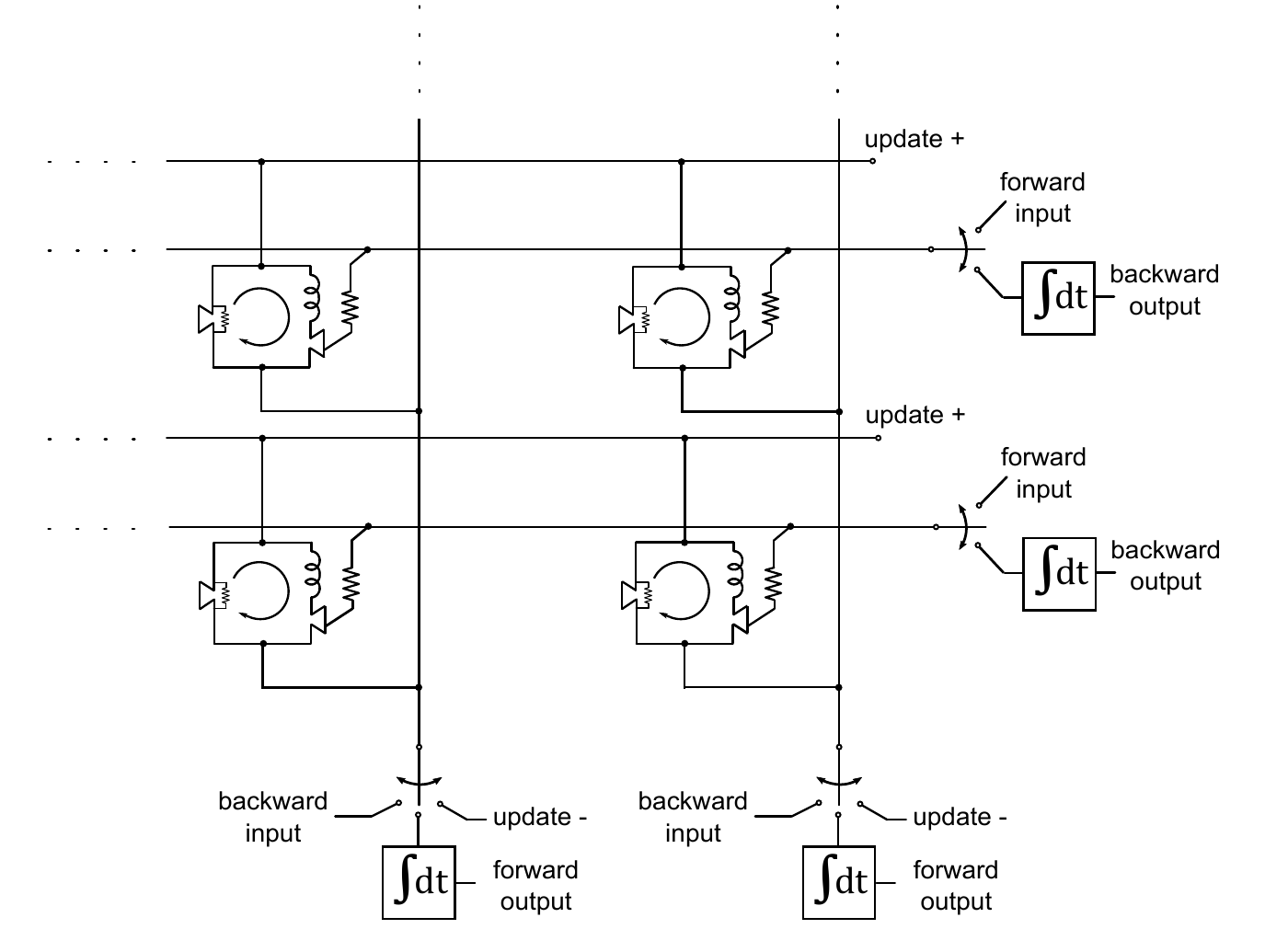}
	\caption{\textit{\textbf{Circuit schematic of a section of the crossbar array and connection schemes under different operations:}} Forward pass involves application of inputs to the rows and the integration of the product at the ends of columns. The backward pass is performed similarly, using the transpose of the matrix. Update pulses are sent from both rows and columns from separate lines which include Josephson Transmission Lines (JTLs) to propagate pulses.}
	\label{fig:Crossbar}
\end{figure}

The forward pass involves computing the product of the input vector with the weight matrix. As explained in Sec.\ref{sec:multiplication}, the transfer function of the unit cell for a given state and output is proportional to $W^2x^{-1}$. This transfer function can be used as the "multiplication" in DNN training with minor algorithmic changes. The division term can be removed by representing inputs as $x^{-1}$ instead of $x$. On the other hand, the $W^2$ term leads to a weight dependent learning rate for each cell. As will be shown in Sec.\ref{sec:emulator}, this non-linear transfer function is suitable for the purposes of the DNN training application.

Calculation of the backward pass can be done by employing the same method, only with the difference of inverted input signs for convention purposes and interchanged input/output terminals to represent the transpose of the weight matrix (Fig.\ref{fig:Crossbar}). 

Updating of the cell follows the stochastic update scheme with the application of pulses from rows and columns with opposite signs. Presence of both pulses provides sufficient current to switch the shunted constriction into its normal state. This switching shuttles an incremental current into, or out of, the loop depending on the input polarity as explained in \ref{sec:SNIPE}. Individual pulses are thresholded by the switching current of the constriction and do not lead to any change. This behavior satisfies the AND operation requirement of the method for implicit and all-parallel calculation of the update \cite{gokmen2016}.

The peripheral circuitry for this crossbar can be built using readily available technologies such as cryo-compatible CMOS and Josephson Junctions (JJs). The elements required for a full-scale integration includes Josephson transmission lines (JTLs), integrators, analog to digital converters (ADCs), digital to analog converters (DACs), and non-linear function evaluators (e.g. CPU or ASIC). The architecture can then be used to accelerate training of Restricted Boltzmann Machines (RBMs), fully connected neural networks\cite{gokmen2016}, convolutional neural networks (CNNs) \cite{gokmen2017}, recurrent neural networks (RNNs), and long-short term memory (LSTM) Networks \cite{gokmen2018training}.

\section{Results}
\label{sec:results}

In this section, we provide simulation and experimental results for the operation of the superconducting nanowire-based cross-point element. Design and fabrication details of the unit cell will be explained. Finally, to demonstrate the applicability of this approach, we have emulated training of different neural networks using the experimentally characterized device response.

\subsection{Simulation of Incrementally Controllable Symmetric States}

\begin{figure}[h!]
	\centering
	\includegraphics[width = \linewidth]{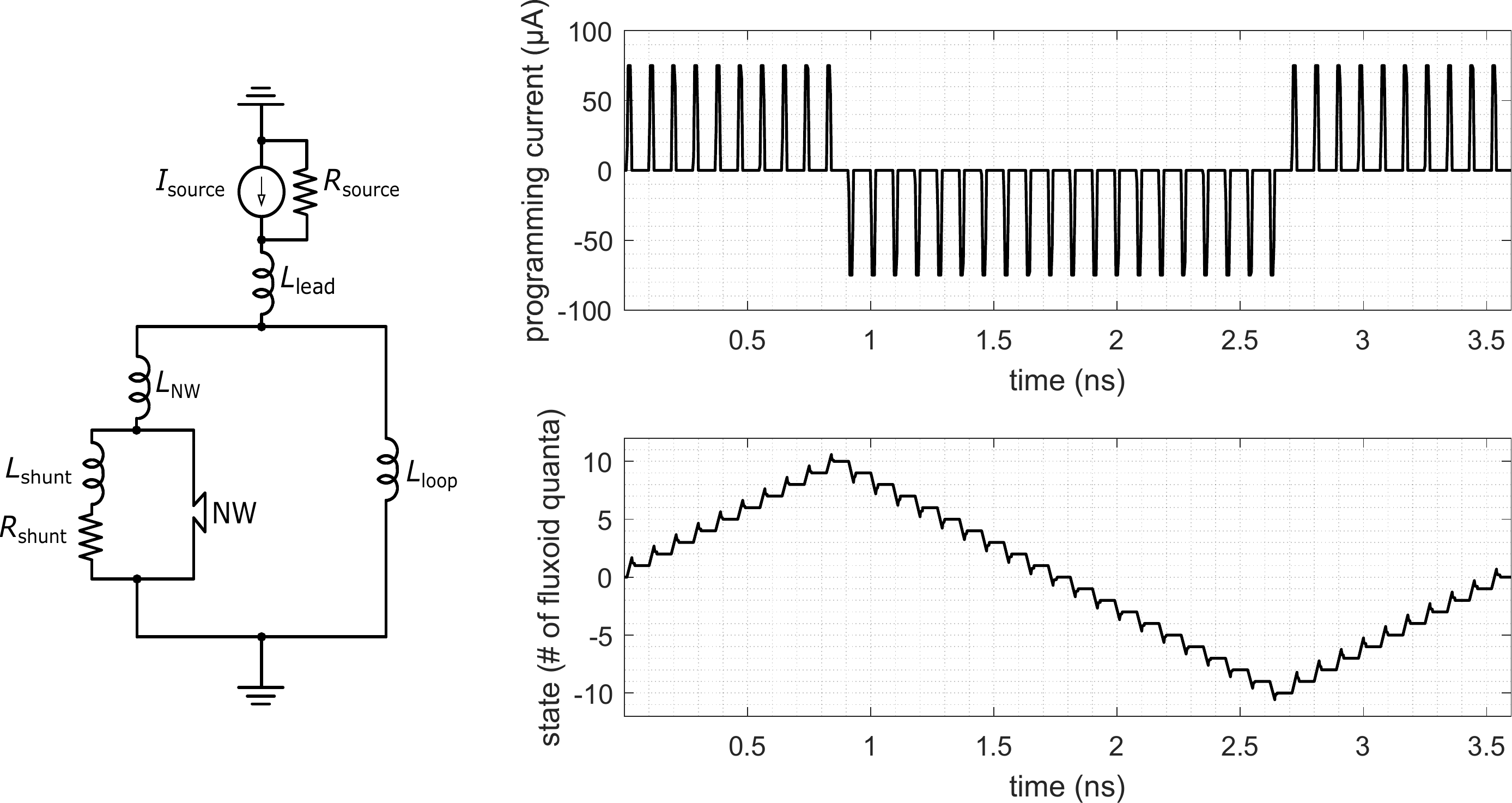}
	\caption{\textit{\textbf{Electro-thermal circuit simulations for the unit cell under incremental programming mode:}} Circuit schematic of the unit cell (left) and simulation results showing incremental state control (right). Pulses of same height (for each polarity) were applied to the cell. In this simulation, the amplitude of the programming pulse was three times the switching current of the constriction and pulse-widths were \SI{15}{\pico\second}. The state of the cell was controlled incrementally, with SFQ equivalent discrete step sizes, in a perfectly symmetric way. Circuit parameters used for this simulation were as following: $R_\textrm{source} = \SI{10}{\kilo\ohm}$, $L_\textrm{lead} = \SI{1}{\nano\henry}$, $L_\textrm{NW} = \SI{150}{\pico\henry}$, $L_\textrm{shunt} = \SI{5}{\pico\henry}$, $R_\textrm{shunt} = \SI{50}{\ohm}$, $I_\textrm{NW, switch} = \SI{25}{\micro\ampere}$, $L_\textrm{loop} = \SI{10}{\nano\henry}$.}
	\label{fig:simulations1}
\end{figure}

The validation of the approach described in this work is first conducted via device level simulations. The circuit simulator used in this work is based on electrothermal dynamics described in Ref.\cite{kerman2009electrothermal}. Details of this software can be found in Sec.\ref{sec:simulator}. The circuit schematic used to represent the cross-point devices is given in Fig.\ref{fig:simulations1}\footnote{The yTron on the right branch is simply represented as an inductor since under normal operation it should never switch.}.

From our simulations, we have determined that the cell can be programmed incrementally through the application of narrow pulses. However, the delivery of such fast pulses ($10-\SI{100}{\pico\second}$) to the cross-point cells could be achieved by using active transmission lines, such as Josephson Transmission Lines (JTL). The use of JTLs to transmit such short pulses is a common practice in the field but for fabrication simplicity, it was not pursued in this work. Therefore, all experimental demonstrations of multi-level programming performed in this work were done with longer pulses ($\ge\SI{10}{\nano\second}$) of variable height (i.e. the devices are programmed in set-mode instead of increment-mode).

\begin{figure}[ht]
	\centering
	\includegraphics[width = 0.8\linewidth]{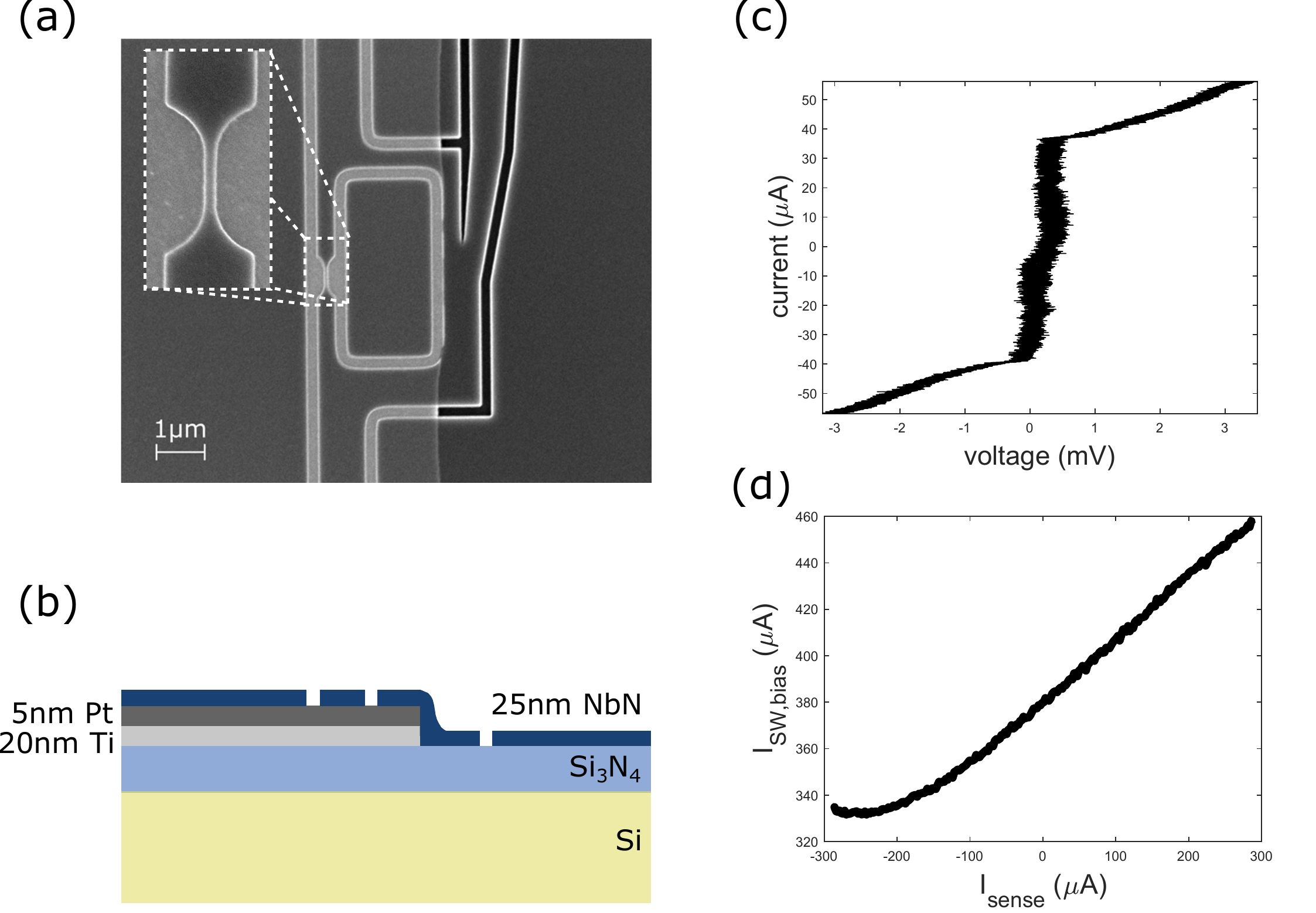}
	\caption{\textit{\textbf{Characterizations of the fabricated unit cell and its subcomponents: }} \textbf{a)} Scanning electron micrograph of the unit cell with a constriction width of \SI{80}{\nano\meter}. The brighter background is a manifestation of the presence of the Pt/Ti shunting layer beneath the constriction side. \textbf{b)} Material stack used for the fabrication of the devices. \textbf{c)} Experimental I-V curve for a stand-alone shunted nanowire. The absence of hysteresis indicates the shunting is effective. Precise measurement of the switching current is performed with a lock-in amplifier and the graph is adjusted accordingly. \textbf{d)} Experimental characterization of a stand-alone yTron that shows the modulation of the switching current of the bias arm as a function of the current flowing in its sense arm.}
	\label{fig:SNIPE}
\end{figure}

\subsection{Design and Fabrication of Unit Cell}
Increasing the number of programmable states for the unit cell requires fine control over the current shuttled per switching event. In our previous work (Ref. \cite{toomey2018bridging}), we showed that this behavior can be obtained by lowering the shunt branch impedance ($R_\textrm{shunt}$, $L_\textrm{shunt}$). Furthermore, simulated results indicate that the increased thermal sinking of the active layer improves the shunting characteristics by providing additional damping over the thermal response (see Sec.\ref{sec:fab}). Considering that at cryogenic temperatures thermal conductivity, in general, follows electrical conductivity, the design can be improved by placing the metal layer directly in contact with the constriction (i.e. \textit{in situ} shunting). This method reduces $L_\textrm{shunt}$ while also functioning as a heat sink for the constriction. Moreover, when a shunted nanowire switches, the voltage response is reduced due to the lower effective normal resistance ($R_\textrm{NW} \parallelsum R_\textrm{shunt}$ ). While this effect is desirable for the constriction; it makes distinguishing the switched and superconducting state of the yTron difficult. For this reason, the yTron's arms were not shunted.

The device mentioned above was fabricated with $\SI{25}{\nano\meter}$ thin film NbN on a metal shunting stack of $\SI{20}{\nano\meter}$ Ti capped with $\SI{5}{\nano\meter}$ Pt to protect the Ti layer (see Sec.\ref{sec:fab} for details). Fig.\ref{fig:SNIPE}a shows a scanning electron micrograph of a cross-point element, where the presence of the metal layer can be seen as a vertical change in contrast. The loop inductance of the unit cell shown in Fig.\ref{fig:SNIPE}a is estimated to be $\approx \SI{527}{\pico\henry}$, which makes the circulating current quantized in $\SI{3.9}{\micro\A}$ steps. Another identical shunted constriction fabricated on this chip was measured to have a switching current of $\approx \SI{36.65}{\micro\A}$. Therefore, under single-flux level control (i.e. each switching event adds or removes a single-flux-quantum to/from the loop), the device is expected to have $\approx$ 19 states ($2I_\textrm{SW}/(L_\textrm{loop}\Phi_\circ)$).
\newpage
\subsection{Experimental Characterization of Multi-Level Programming and Analog Multiplication}

\begin{figure}[h!]
	\centering
	\includegraphics[width =  \linewidth]{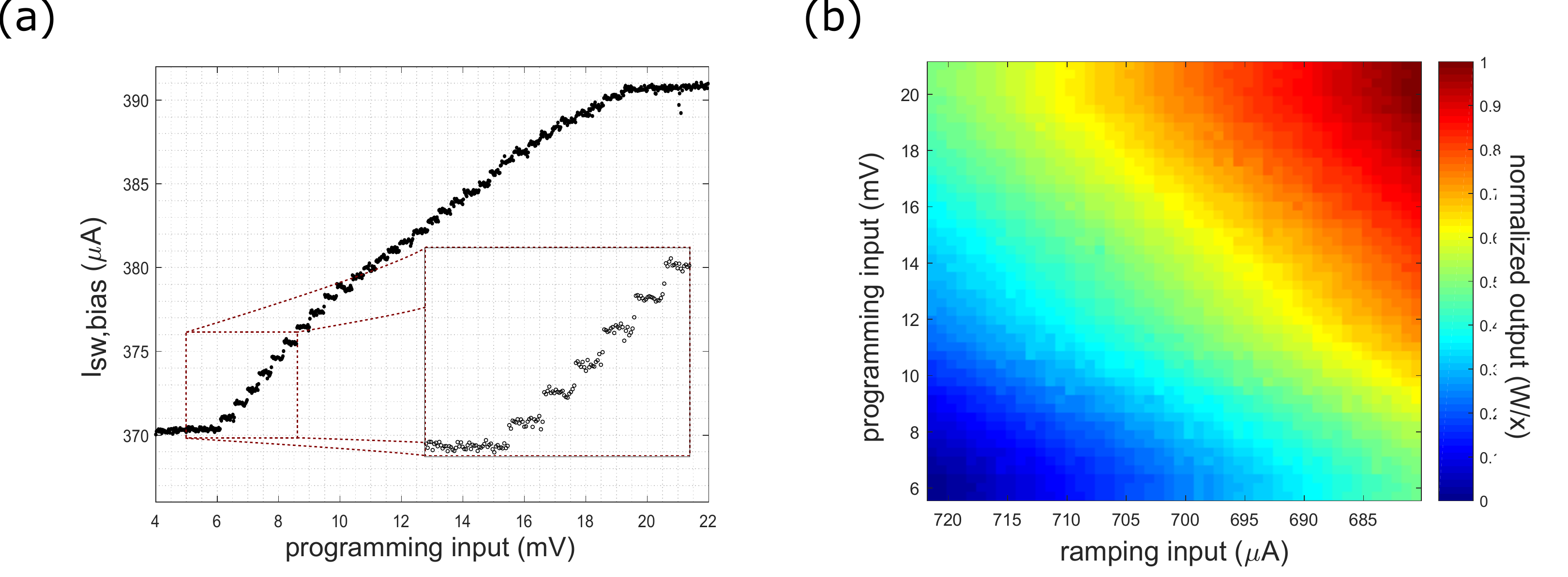}
	\caption{\textit{\textbf{State diagram and analog multiplication results for the fabricated unit cell: }} \textbf{a)}Switching current of the yTron's bias arm (state of the cell) as a function of the programming input. First, a programming input was sent to the device to set the state of the cell. Then, the readout was performed by sending a ramping input to the yTron's bias arm and recording current at which it switched. This operation was repeated for a range of programming inputs to obtain the state diagram of the unit cell. We have observed 33 discrete levels which can be thought of as a 5-bit memory. \textbf{b)} Analog multiplication results for the unit cell. Similar to the generation of the state diagram, the cell was first set to a certain state using 50 different programming inputs. Then, 50 different ramping inputs were used to perform the analog multiplication operation described in Sec.\ref{sec:multiplication}, by recording the time (current) at which yTron's bias arm switches. Results are presented as a colormap where the normalized output of the operation is represented with the color, versus cross-swept programming and ramping inputs.}
	\label{fig:experimental_merged}
\end{figure}

The state diagram for the unit cell in Fig.\ref{fig:SNIPE}a is shown in  Fig.\ref{fig:experimental_merged}a. It can be seen that 33 discrete states could be resolved with this device. This number is greater than the calculated number of states, which strongly suggests that we are operating in the SFQ limit. We attribute this discrepancy between the estimated and realized number of states to an underestimate of loop inductance. Specifically, the superconducting region that is in direct contact with the metal is expected to have a higher kinetic inductivity as a result of an effective thinning due to suppression by the shunt. Therefore we expect to have a finer quantization and a greater number of states as observed.

We then demonstrated the device operating in the proposed multiplication scheme. Due to the oscilloscope having finer time resolution than voltage resolution, we detected the switching time ($\propto W/x$) instead of the integrated area (Fig.\ref{fig:Operations}, $\propto W^2/x$)\footnote{In a large-scale implementation, this operation would be achieved using on-chip integrators, with the method mentioned in Sec.\ref{sec:multiplication}}. Fig.\ref{fig:experimental_merged}b shows the normalized output as a function of input level and cell state (programming level). It can be seen that the output is proportional to the product of input and weight with the expected characteristics. Due to the existing of the offsets in the multiplicands, the nonlinearity of multiplication appears less pronounced. A differential readout (i.e. with a baseline device). It is common in crosspoint architectures to use a differential reading between the main array and a reference array \cite{gokmen2016}. We expect the same approach to be used in this application, as this provides negative weights and eliminates the offset. Furthermore, increasing the sensitivity of the yTrons would further improve the devices' characteristics.

\subsection{Emulation of DNN Training Using Characterized Device Response}
\label{sec:emulator}

Following the experimental characterizations, we have implemented an emulator to investigate how well these devices would perform in a crossbar architecture, being used to train actual DNNs. For this purpose, we have selected two common benchmark networks and trained them on an image classification task of handwritten digits (MNIST dataset). The details of these networks and the method used to emulate the crossbar architectures can be found in Refs.\cite{gokmen2016,gokmen2017}.

\begin{figure}[h!]
	\centering
	\includegraphics[width =  \linewidth]{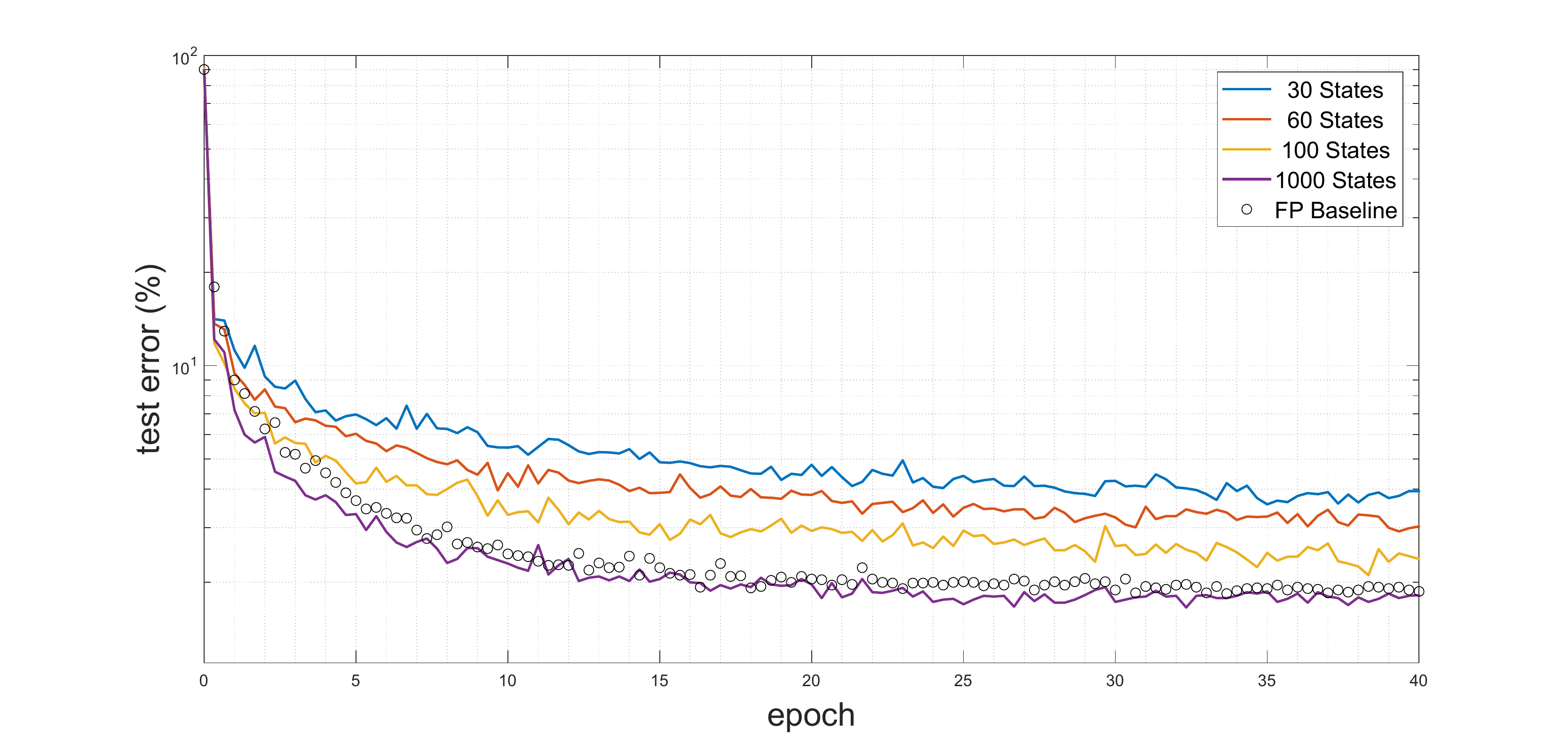}
	\caption{\textit{\textbf{Emulated training curves of networks on the MNIST dataset, trained by the crossbar architectures with the characterized device properties:}} Graphs show classification error on the test set, with respect to increased training. Each epoch consists of updating the network for all 60000 images in the MNIST training set. Fully-connected neural network is composed of four layers of 784, 256, 128, and 10 neurons, respectively \cite{gokmen2016}.}
	\label{fig:network_responses}
\end{figure}

We have modeled our devices as perfectly symmetric switching cross-point elements that have quadratic state diagrams to take into account the $W^2x^{-1}$ behavior observed in multiplication. These emulations included analog noise, signal bounds (saturation in operational amplifiers), ADC/DAC resolutions and stochastic update scheme, but neglected device-to-device variations (see Sec.\ref{sec:emulator_supplementary}). In order to see how the performance depends on the number of programmable states, each network is trained for four cases of 30, 60, 100, and 1000 states. Fig.\ref{fig:network_responses} shows the learning curve of the fully connected neural network on the MNIST dataset \cite{gokmen2016}. 
  
The results in Fig.\ref{fig:network_responses} show that even with the current capacity of $\approx$30 states per device, 96.74\% testing accuracy can be achieved with the fully connected neural network (with respect to the 98.2\% baseline, trained with conventional digital processors). Furthermore, this performance degradation diminished when the number of states was increased to 1000. Similarly, the convolutional neural network provided 98.86\% testing accuracy (with respect to the 99.2\% baseline performance). However, this performance required 1000 states from the device, as suggested by Ref.\cite{gokmen2017} for linear memristive devices. We have not observed any additional performance degradation due to the non-linear transfer characteristics of our devices. These results signify that the unit cell shown in this work is a promising candidate to realize the cross-point elements for acceleration of DNN training.

\section{Discussion and Conclusions}

In this work, we have designed, fabricated, and tested a superconducting nanowire-based inductive processing element as a cross-point device (Fig.\ref{fig:cartoon}). Programming of these devices at the single flux quantum (SFQ) limit allowed resolution of 33 discrete, perfectly non-volatile states (Fig.\ref{fig:experimental_merged}). We have proposed and demonstrated a method to perform analog multiplication of the input with the state of these devices (Fig.\ref{fig:Operations}). Furthermore, using our electro-thermal circuit simulator, we presented the requirements necessary to use the cell in the incremental programming mode which leads to perfectly symmetric switching states (Fig.\ref{fig:simulations1}). Operation of these devices in a crossbar is described for full DNN training that implements stochastic gradient descent using the back-propagation algorithm for calculation of the gradients (Fig.\ref{fig:Crossbar}). The concept is further validated using an emulator which predicts classification errors close to floating point processor baselines, using the characterized device responses (Fig.\ref{fig:network_responses}). 

The former system-level analysis in Ref.\cite{gokmen2016,gokmen2017} has shown that having symmetric switching cross-point elements is crucial for building crossbar based DNN training accelerators without causing significant performance degradation. The devices made in this work provides this feature, unlike any other non-volatile memory alternatives. The current number of states was found to be sufficient to train the fully connected neural network as shown in Fig.\ref{fig:network_responses} however, it needs to be increased to address more complex paradigms (such as convolutional neural networks or larger scale DNNs). The number of programmable levels can be increased by optimizing the readout mechanism and the device layout. While addressing 100 states with these devices is realistic in the near future, another order of magnitude might become challenging without increasing the device size. 

The new analog 'multiplication' process we have implemented is suitable for the purposes of the application, although it is not exactly multiplication. We have shown that the non-linearity of the multiplication scheme was tolerable by the emulated network training performances (Fig.\ref{fig:network_responses}). These results are in good agreement with the findings of prior studies \cite{gokmen2016,gokmen2017,gokmen2018training}. Moreover, such an integration-time-modulation-based analog multiplication can potentially be used in other existing memristive device families.

The main limitation we see before a large-scale implementation is the routing of the signals, particularly in the update cycle. As the crossbar in this approach is operated with currents, the signaling paths require directional elements to propagate the pulses correctly. On the other hand, forward and backward pass cycles do not suffer from this problem as they can be controlled using voltage pulses, which are then locally converted to current pulses at each yTron. A superconducting nanowire-Josephson junction hybrid architecture can potentially overcome these challenges to provide a large-scale (e.g. 1000$\times$1000) crossbar array. Alternatively, optical pulses might be used to replace one of the update signals, while the other one functions as a bias input that determines the direction of the update. 

The devices presented in this work are designed in a particular way such that their weak points can be tolerated by the application and vice versa. Future work will include scaling up the unit cells to increase memory capacity and programming the device state incrementally. Ultimately, the concept presented in this work shows promising characteristics and has the potential to realize acceleration of DNN training without introducing network performance degradation.
\newpage

\section{Acknowledgments}

This work was supported in part by the Intelligence Advanced Research Projects Activity (IARPA) and the U.S. Army Research Office under contract number W911NF-14-C-0089. The content of the information does not necessarily reflect the position or the policy of the Government, and no official endorsement should be inferred. The authors gratefully acknowledge further support from Intel Corporation.

\section{Supplementary Materials}

\subsection{Electrothermal Circuit Simulator}
\label{sec:simulator}

We have built a circuit simulator in the MATLAB environment. In order to model the behavior of the superconducting nanowires, we have used the hot-spot formation and decay dynamics discussed in Ref.\cite{kerman2009electrothermal}. A SPICE implementation of the same model is also existent as reported in Ref.\cite{berggren2018superconducting}. The simulator used in this work has a built-in dynamic time stepping algorithm to provide high accuracy and performance. Simulator inputs the circuit netlist, input waveforms and physical parameters that define the system and outputs the node voltages and branch currents by solving the system for the thermal and electrical equations. Further details and other applications of this simulator can be found in Ref.\cite{toomey2018bridging}.

Superconducting nanowire-based devices (as well as the conventional electrical circuit components such as resistors, inductors etc.) do not require capturing the phase information to describe their operation. In other words, they are not governed by quantum coherent transport and therefore the evolution of the superconducting phase across these devices is neglected. Instead, the quantization of flux in the superconducting loop is enforced when the nodal and branch quantities reach steady state (following their computation in the absence of such a constraint).

\begin{figure}[h!]
	\centering
	\includegraphics[width = \linewidth]{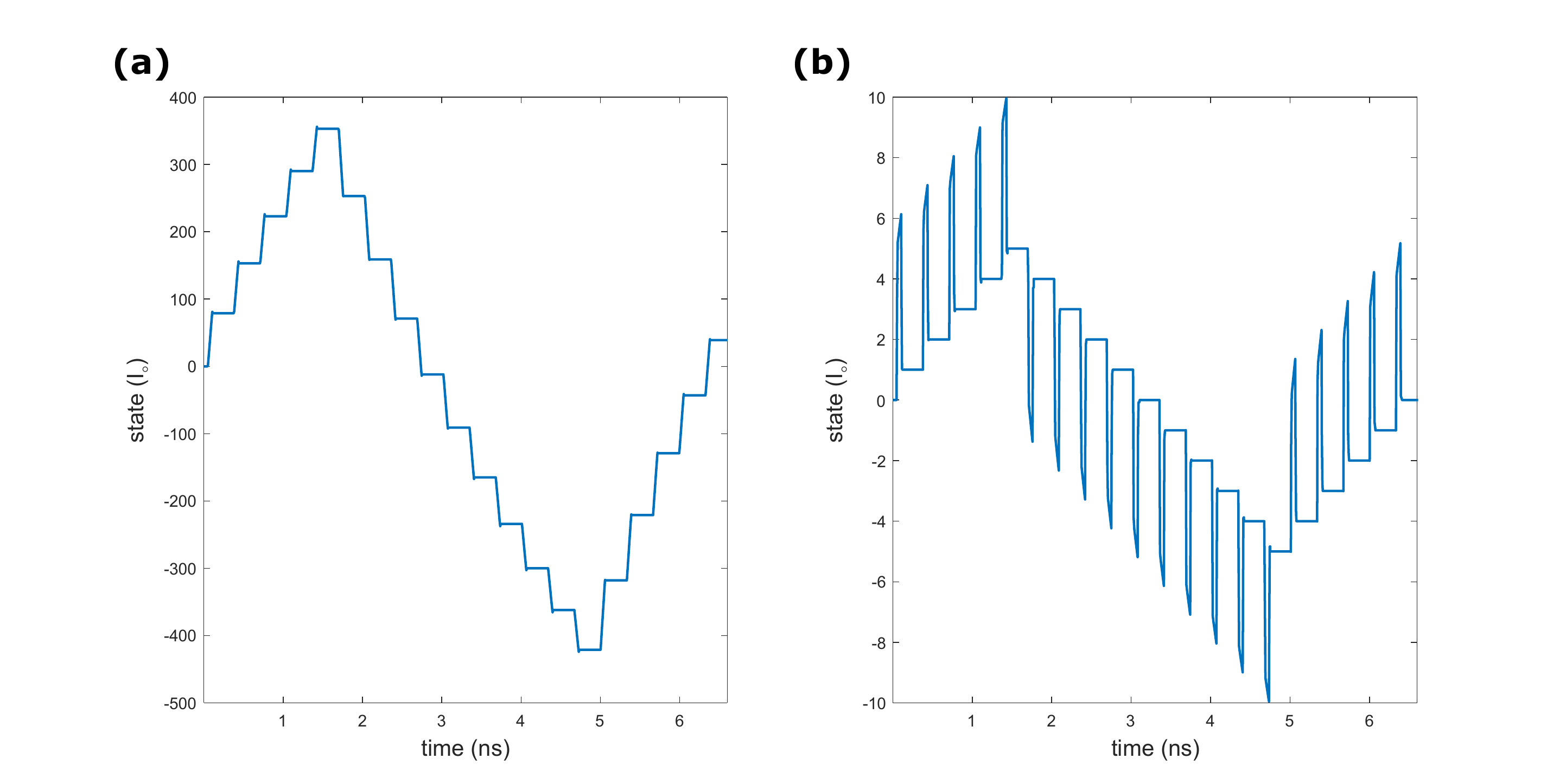}
	\caption{\textit{\textbf{Simulation results for two devices with different thermal conductivities and the same $R_\textrm{s}$ value:}} \textbf{(a)} Regular electrical shunting with a relatively large shunt resistor ($R_\textrm{s}$ = \SI{50}{\ohm}). \textbf{(b)} Optimized, electrical and thermal, shunting with the same shunt resistor but 100 times higher thermal conductivity. Optimized shunting method can achieve SFQ level control even with $R_\textrm{s}$ = \SI{50}{\ohm}.}
	\label{fig:thermal_shunt}
\end{figure}

Finally, the simulated results for the effect of thermal sinking can be found in Fig.\ref{fig:thermal_shunt}. Results shown here suggest that the increased thermal conductivity of the substrate can potentially improve the effectiveness of the shunt resistor.

At cryogenic temperatures, thermal conductivity follows electrical conductivity (due to the absence of crystal motion caused phononic conductivity). For the conventional substrates, such as Si, $\mathrm{Si_2}$, $\mathrm{Si_3N_4}$, $\mathrm{Al_2O_3}$\footnote{Thermal conductivity of $\mathrm{Al_2O_3}$ is considerably high at room temperature due to the stiff crystal structure and high speed of sound of the material. However, this term becomes negligible when the material is cooled down to cryogenic temperatures since the lattice movements disappear.} etc. thermal conductivity is very low, as all of them are electrical insulators. 
However, when the constriction is patterned above a metal island, this layer can work as a heat sink (i.e. a substrate with high thermal conductivity) and an electrical shunt at the same time.

It must also be noted that this modified shunting method brings some other design concerns such as proximity effect and heating effect of the metal, both of which can potentially suppress the superconductivity of the active layer. We have used a thick NbN layer to avoid such a problem. Considering that the electrothermal picture of these systems is rather complicated, a further extensive analysis will be required to validate the effect shown in Fig.\ref{fig:thermal_shunt}. Nonetheless, the shunting method presented method is proven to be effective in controlling the nanowire behavior as shown in Fig.\ref{fig:experimental_merged}a.

\subsection{Fabrication of the Unit Cell}
\label{sec:fab}
We fabricated these devices on \SI{100}{\nano\meter}  $\mathrm{Si_3N_4}$ on Si substrate. The metal layer was patterned by a liftoff process (for the fabrication of the shunting islands). For this purpose, we have used a double layer stack of PMGI-SF9 and S1813, exposed by using Heidelberg A101 direct writer with \SI{7}{\milli\watt} beam at 25\% and developed in CD-26. Liftoff was performed by sonicating the chip in acetone for \SI{180}{\second}. The active NbN layer was then sputtered using magnetron sputtering at room temperature, following the process described in Ref.\cite{dane2017bias}. The sheet resistance of the film was calculated using a four-point probe measurement. Patterning of the active layer was performed using electron beam lithography. We used ZEP520A as the resist and back-scattered electron detector for the alignment with the former layer. Exposure was done using Elionix FLS-125, at \SI{500}{\micro\coulomb/{\centi\meter\squared}} dose with \SI{500}{\pico\ampere} beam current for small features (devices) and \SI{40}{\nano \ampere} for larger features. The resist was developed in o-Xylene at \SI{5}{\celsius} for \SI{90}{\second} followed by \SI{30}{\second} dip in isopropanol and \textrm{$N_2$} gun dry. NbN was etched with reactive ion etching using $\mathrm{CF_4}$ at \SI{50}{\watt}, for a total period of \SI{360}{\second} partitioned in 3 equal steps. Excess resist was stripped in n-methyl-2-pyrrolidone (NMP) at \SI{70}{\celsius} for \SI{1}{\hour}. A Zeiss Orion SEM was used to image the devices at \SI{5}{\kilo\volt} acceleration voltage at \SI{4.8}{\milli\meter} working distance using the IL detector with \SI{30}{\micro\meter} aperture width.

Device shown in Fig.\ref{fig:SNIPE} was originally designed to have \SI{100}{\nano\meter} constriction width and \SI{400}{\nano\meter} wise yTron arms. We have observed that the fabrication yielded in devices that are $\approx$\SI{10}{\nano\meter} shorter/narrower in all directions. Therefore, we reflected this back to the layout and calculated that the fabricated device had 35.3 squares. This calculation was performed by using a basic environment in COMSOL Multiphysics software. In order to compute the kinetic inductivity we have used the formula $L_{kinetic} = 1.38\frac{R_{sheet}}{T_c}$. We have measured on a sister chip that the sheet resistance and the critical temperature of the active NbN layer were \SI{97.2}{\ohm}$/\Box$ and \SI{9.0}{\kelvin} respectively providing a kinetic inductivity of \SI{14.9}{\pico\henry}$/\Box$. Therefore, we have concluded that the total loop inductance is \SI{526.6}{\pico\henry} leading to quantization of circulating current in \SI{3.93}{\micro\ampere} steps. Note that this is an underestimation considering that the shunted region is expected to have a higher kinetic inductivity.

\subsection{Experimental Setup and Processing of Raw Data}

Experimental characterizations of these devices were made under liquid helium immersion conditions using LeCroy Waverunner 620Zi \SI{2}{\giga\hertz}  oscilloscope, and Agilent 33600A Trueform Series arbitrary waveform generator (AWG). Waveforms were programmed using MATLAB, which controlled the AWG and also acquired the output waveform from the oscilloscope. We have used DC-\SI{4.2}{\giga\hertz} splitters (Mini-Circuits ZFRSC-42) to divide the signals (programming pulses and readout ramps) between the devices and the oscilloscope. Although the device operations are all current controlled, we have chosen to report all of the results in voltage levels. This choice was made to avoid any misrepresentation since we were not able to accurately measure the current values going into the device ports. 

Similarly, we have reported the read trigger times of the yTron bias arm, instead of the current equivalents of these values. These measurements we conducted by first getting the voltage waveform appearing on the bias arm of the yTron. Then, using an arbitrary thresholding level, the switching time of the bias arm was registered through the built-in functions of the LeCroy Waverunner 620Zi. We note that a similar measurement could have also been implemented using the integration functions of the oscilloscope. This method would have been closer to the technique we have proposed for the multiplication as well. However, it was realized that the timing resolution of the instrument was superior to that of the amplitude resolution. Similar measurement techniques and details can be found in Ref.\cite{toomey2018bridging}.

Each data-point shown in Fig.\ref{fig:experimental_merged}a is an average of 50 consecutive readouts, following a single programming input. When the dynamic range of the states is normalized in between 0 to 1, the average standard deviation of read trigger level was found to be $1.60\times10^{-4}$. For practical purposes, we chose not to plot the results in error-bars, as the standard deviation is very small. This result also suggests that the readout could have been performed at a single run as well. 

The multiplication results shown in Fig.\ref{fig:experimental_merged}b are obtained very similarly to that of the state diagram mentioned above. We have first adjusted the ranges of the programming input and the ramping input in a way, such that the diagonal values are approximately the same. This choice was made in order to clearly represent the effects of each input on the output. Between each ramping input, the state of the device was reset with a high negative pulse. Following the former results, showing low standard variation across successive readouts, the number of readouts per point is reduced to 10 in this experiment. We have observed for a given ramping input, the average standard deviation (following the same normalization procedure described above) to be $1.04\times10^{-4}$. 

On the other hand, it can be seen that the sensitivity to the ramping input is still higher than that to the programming input. This behavior manifests itself in the constant output contours shown in Fig.\ref{fig:experimental_merged}b, where the curvatures are less than expected. One way to analyze this issue is to consider it as a limited dynamic range problem as any multiplication contour-map would appear linear when the axes are not balanced. Therefore, improved yTron designs might be able to address this issue. Nonetheless, for the application of DNN training, similar to the quadratic behavior, this imperfection can be tolerated by the algorithm.

\subsection{Reproducibility and Additional Results}
\label{sec:additional}
In addition to the device reported in the main body, we have fabricated devices with different yTron geometries and loop sizes. We have observed the yTron behavior varied even depending on whether the device was connected to a loop geometry or not. This behavior can be seen between the results reported in Fig.\ref{fig:SNIPE}d and Fig.\ref{fig:experimental_merged}, where the former has a very linear response in the operating regime while the latter has a second deflection point. We suspect that this characteristic can be attributed to switching different spots on the bias arm under different conditions. However, we have not found conclusive experimental evidence to diagnose the exact reasoning behind this phenomenon. Nonetheless, for the purposes of demonstrating the operation of our cross-point elements, all of the yTrons we have tested proved to be sensitive to the changes in the circulating current (i.e. operational). It remains as an open question how to optimize yTron geometry by means of widths of the arms, connection angle, and distance from the corners of the main loop.

\begin{figure}[ht]
	\centering
	\includegraphics[width = 0.8\linewidth]{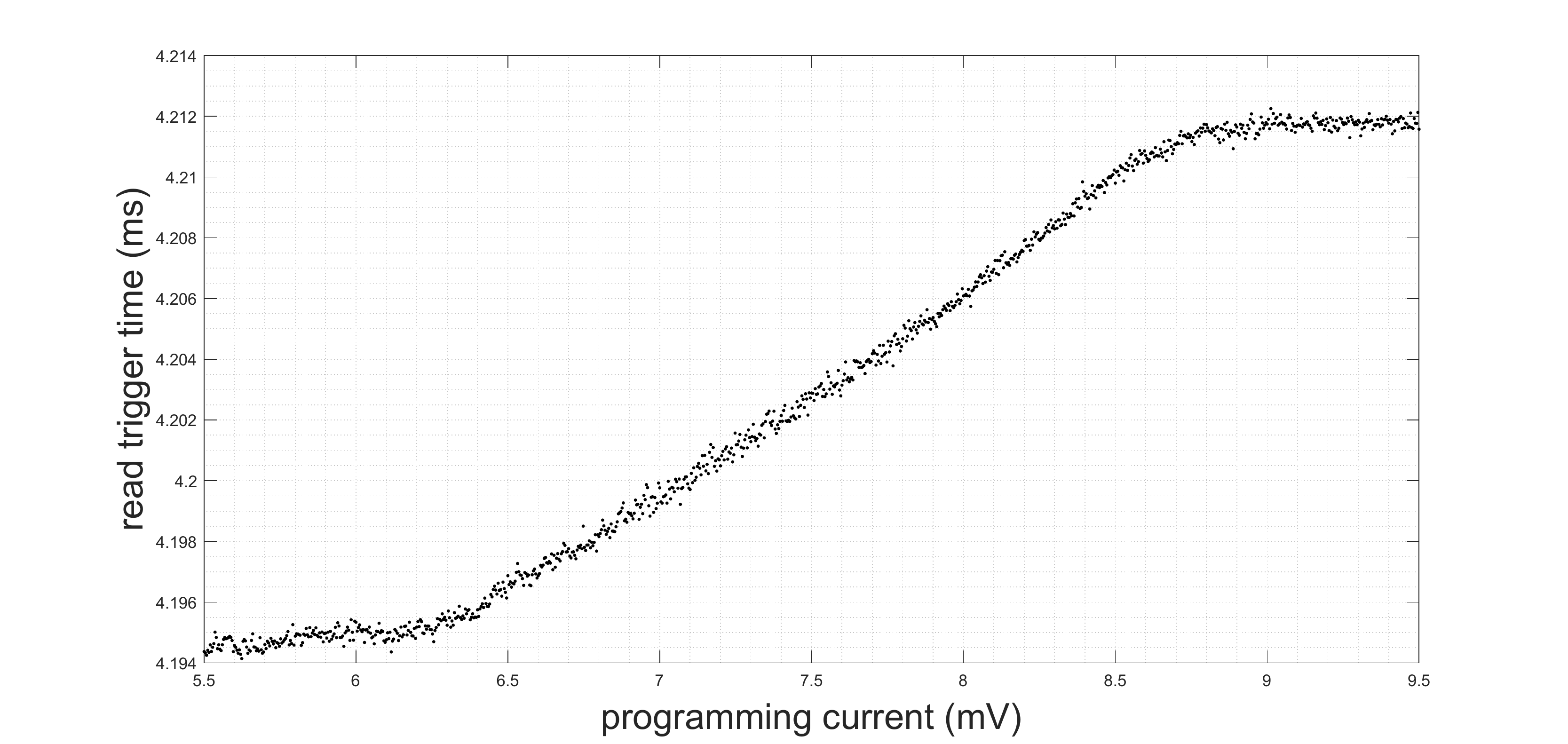}
	\caption{\textit{\textbf{State diagram for the larger loop based cell: }} Analogous results to the ones shown in Fig.\ref{fig:experimental_merged}. Loop inductance is \SI{3.94}{\nano\henry}, leading to quantization of current with \SI{525.16}{\nano\ampere} steps. A continuum of states is observed due to the resolution limit of the yTron and the setup.}
	\label{fig:long_loop}
\end{figure}

Furthermore, we have also experimented with a loop that is larger than the device we report in the main body. Following the similar calculations described above, this loop was found to have a total inductance of \SI{3.94}{\nano\henry}, leading to quantization of current with \SI{525.16}{\nano\ampere} steps. Fig.\ref{fig:long_loop} shows the state diagram for this device. This device shows a continuum of 'states', however, due to the absence of observable quantization effects, defining them as states is not practical. Analyses of these result suggest that the number of states is limited by the readout resolution instead of the programming capability of the shunted constriction. Each data-point shown in this graph is an average of 50 consecutive readouts without any rewriting similar to the one explained above in the experimental setup section. When the dynamic range is normalized in between 0 to 1, the average standard deviation for each programming input was found to be $1.03\times10^{-3}$ which is approximately 6.4 times the average deviation observed for the small device.

\subsection{DNN Emulator}
\label{sec:emulator_supplementary}
The DNN emulator used in this work is a direct implementation of the software developed and used in Refs.\cite{gokmen2016,gokmen2017,gokmen2018training}. For comparison with the studies conducted in Refs.\cite{gokmen2016,gokmen2017,gokmen2018training}, Table \ref{table:network parameters} is provided below. Noise and bound management techniques discussed in Ref.\cite{gokmen2017} are used in all emulations as well.

\begin{table}[h!]
	\centering
	\small
	\begin{tabular}{llllllll}
		\hline
		\multicolumn{1}{c}{Parameter} &Mini-batch Size &Learning Rate &Bit-length &Noise &Signal Bound &DAC &ADC\\ \hline
		Value              &1 &0,01 &10 &0.06 &12 &5-bit &9-bit \\\hline
	\end{tabular}
	\caption{Hyperparameter set used in the DNN emulator. For detailed explanation of the system see Ref. \cite{gokmen2016}.}. 
    \label{table:network parameters}
\end{table}
 
Furthermore, when the device is set to the highest state, additional incremental pulses do not lead to any change, which can be viewed as a saturated behavior. All devices were assumed to be the same due to the absence of such statistical information to us at the moment. Due to the quantum-mechanically dictated quantization levels, we do not expect radical variations in the number of states. However, the individual devices might have different weight coefficients which is neglected in these simulations. A detailed variation sensitivity analysis can be found in Ref.\cite{gokmen2016}.

\newpage
\begin{singlespace}
\bibliography{main}

\newcommand{\noopsort}[1]{} \newcommand{\printfirst}[2]{#1}
  \newcommand{\singleletter}[1]{#1} \newcommand{\switchargs}[2]{#2#1}
\begin{thebibliography}{10}

\bibitem{lecun2015}
Y.~LeCun, Y.~Bengio, and G.~Hinton, ``Deep learning,'' {\em nature}, vol.~521,
  no.~7553, p.~436, 2015.

\bibitem{krizhevsky2012}
A.~Krizhevsky, I.~Sutskever, and G.~E. Hinton, ``Imagenet classification with
  deep convolutional neural networks,'' in {\em Advances in neural information
  processing systems}, pp.~1097--1105, 2012.

\bibitem{collobert2011}
R.~Collobert, J.~Weston, L.~Bottou, M.~Karlen, K.~Kavukcuoglu, and P.~Kuksa,
  ``Natural language processing (almost) from scratch,'' {\em Journal of
  Machine Learning Research}, vol.~12, no.~Aug, pp.~2493--2537, 2011.

\bibitem{najafabadi2015deep}
M.~M. Najafabadi, F.~Villanustre, T.~M. Khoshgoftaar, N.~Seliya, R.~Wald, and
  E.~Muharemagic, ``Deep learning applications and challenges in big data
  analytics,'' {\em Journal of Big Data}, vol.~2, no.~1, p.~1, 2015.

\bibitem{burr2017neuromorphic}
G.~W. Burr, R.~M. Shelby, A.~Sebastian, S.~Kim, S.~Kim, S.~Sidler, K.~Virwani,
  M.~Ishii, P.~Narayanan, A.~Fumarola, {\em et~al.}, ``Neuromorphic computing
  using non-volatile memory,'' {\em Advances in Physics: X}, vol.~2, no.~1,
  pp.~89--124, 2017.

\bibitem{ambrogio2018equivalent}
S.~Ambrogio, P.~Narayanan, H.~Tsai, R.~M. Shelby, I.~Boybat, C.~Nolfo,
  S.~Sidler, M.~Giordano, M.~Bodini, N.~C. Farinha, {\em et~al.},
  ``Equivalent-accuracy accelerated neural-network training using analogue
  memory,'' {\em Nature}, vol.~558, no.~7708, p.~60, 2018.

\bibitem{chi2016prime}
P.~Chi, S.~Li, C.~Xu, T.~Zhang, J.~Zhao, Y.~Liu, Y.~Wang, and Y.~Xie, ``Prime:
  A novel processing-in-memory architecture for neural network computation in
  reram-based main memory,'' in {\em ACM SIGARCH Computer Architecture News},
  vol.~44, pp.~27--39, IEEE Press, 2016.

\bibitem{prezioso2015training}
M.~Prezioso, F.~Merrikh-Bayat, B.~Hoskins, G.~Adam, K.~K. Likharev, and D.~B.
  Strukov, ``Training and operation of an integrated neuromorphic network based
  on metal-oxide memristors,'' {\em Nature}, vol.~521, no.~7550, p.~61, 2015.

\bibitem{kim2011functional}
K.-H. Kim, S.~Gaba, D.~Wheeler, J.~M. Cruz-Albrecht, T.~Hussain, N.~Srinivasa,
  and W.~Lu, ``A functional hybrid memristor crossbar-array/cmos system for
  data storage and neuromorphic applications,'' {\em Nano letters}, vol.~12,
  no.~1, pp.~389--395, 2011.

\bibitem{steinbuch1961lernmatrix}
K.~Steinbuch, ``Die lernmatrix,'' {\em Biological Cybernetics}, vol.~1, no.~1,
  pp.~36--45, 1961.

\bibitem{coppersmith1987matrix}
D.~Coppersmith and S.~Winograd, ``Matrix multiplication via arithmetic
  progressions,'' in {\em Proceedings of the nineteenth annual ACM symposium on
  Theory of computing}, pp.~1--6, ACM, 1987.

\bibitem{gokmen2016}
T.~Gokmen and Y.~Vlasov, ``Acceleration of deep neural network training with
  resistive cross-point devices: design considerations,'' {\em Frontiers in
  neuroscience}, vol.~10, p.~333, 2016.

\bibitem{gokmen2017}
T.~Gokmen, M.~Onen, and W.~Haensch, ``Training deep convolutional neural
  networks with resistive cross-point devices,'' {\em Frontiers in
  neuroscience}, vol.~11, p.~538, 2017.

\bibitem{vincent2010stacked}
P.~Vincent, H.~Larochelle, I.~Lajoie, Y.~Bengio, and P.-A. Manzagol, ``Stacked
  denoising autoencoders: Learning useful representations in a deep network
  with a local denoising criterion,'' {\em Journal of machine learning
  research}, vol.~11, no.~Dec, pp.~3371--3408, 2010.

\bibitem{toomey2018bridging}
E.~Toomey, M.~Onen, M.~Colangelo, B.~Butters, A.~McCaughan, and K.~Berggren,
  ``Bridging the gap between nanowires and josephson junctions: A
  superconducting device based on controlled fluxon transfer,'' {\em Physical
  Review Applied}, vol.~11, no.~3, p.~034006, 2019.

\bibitem{mccaughan2016}
A.~N. McCaughan, N.~S. Abebe, Q.-Y. Zhao, and K.~K. Berggren, ``Using geometry
  to sense current,'' {\em Nano letters}, vol.~16, no.~12, pp.~7626--7631,
  2016.

\bibitem{annunziata2010tunable}
A.~J. Annunziata, D.~F. Santavicca, L.~Frunzio, G.~Catelani, M.~J. Rooks,
  A.~Frydman, and D.~E. Prober, ``Tunable superconducting nanoinductors,'' {\em
  Nanotechnology}, vol.~21, no.~44, p.~445202, 2010.

\bibitem{rumelhart1986learning}
D.~E. Rumelhart, G.~E. Hinton, and R.~J. Williams, ``Learning representations
  by back-propagating errors,'' {\em nature}, vol.~323, no.~6088, p.~533, 1986.

\bibitem{gokmen2018training}
T.~Gokmen, M.~Rasch, and W.~Haensch, ``Training lstm networks with resistive
  cross-point devices,'' {\em arXiv preprint arXiv:1806.00166}, 2018.

\bibitem{kerman2009electrothermal}
A.~J. Kerman, J.~K. Yang, R.~J. Molnar, E.~A. Dauler, and K.~K. Berggren,
  ``Electrothermal feedback in superconducting nanowire single-photon
  detectors,'' {\em Physical review B}, vol.~79, no.~10, p.~100509, 2009.

\bibitem{berggren2018superconducting}
K.~K. Berggren, Q.-Y. Zhao, N.~Abebe, M.~Chen, P.~Ravindran, A.~McCaughan, and
  J.~C. Bardin, ``A superconducting nanowire can be modeled by using spice,''
  {\em Superconductor Science and Technology}, vol.~31, no.~5, p.~055010, 2018.

\bibitem{dane2017bias}
A.~E. Dane, A.~N. McCaughan, D.~Zhu, Q.~Zhao, C.-S. Kim, N.~Calandri,
  A.~Agarwal, F.~Bellei, and K.~K. Berggren, ``Bias sputtered nbn and
  superconducting nanowire devices,'' {\em Applied Physics Letters}, vol.~111,
  no.~12, p.~122601, 2017.

\end{thebibliography}
\bibliographystyle{plain}
\end{singlespace}

\end{document}